\documentclass[showpacs,preprintnumbers,amsmath,amssymb,preprint,aps,pre]{revtex4}
\usepackage{graphicx}
\usepackage{amsfonts}
\begin{document}

\title{Anticipated synchronization and the predict-prevent control method in the FitzHugh-Nagumo model system}
\author{C. Mayol, C.R. Mirasso, R. Toral}
\affiliation{IFISC (Instituto de F\'{\i}sica Interdisciplinar y Sistemas Complejos), CSIC-UIB, Campus UIB, E-07122 Palma de Mallorca, Spain}

\begin{abstract}
We study the synchronization region of two unidirectionally coupled, in a master-slave configuration, FitzHugh-Nagumo systems under the influence of external forcing terms. We observe that anticipated synchronization is robust to the different types of forcings. We then use the predict-prevent control method to suppress unwanted pulses in the master system by using the information of the slave output. We find that this method is more efficient than the direct control method based on the master. Finally, we observe that a perfect matching between the parameters of the master and the slave is not necessary for the control to be efficient. Moreover, this parameter mismatch can, in some cases, improve the control.
\end{abstract}
\date{\today}
\maketitle

\section{Introduction}
Synchronization of nonlinear systems is a common phenomenon studied in physical, chemical and biological systems, among others \cite{pik}. One of the most astonishing cases is the so-called {\sl anticipated synchronization}: when two dynamical systems are unidirectionally coupled in a master-slave  configuration including appropriate delay terms, the slave can predict the trajectory of the master \cite{voss}. Remarkably, it has been proved that this kind of synchronization is stable and robust even in the presence of an external forcing acting upon both systems. Besides the many theoretical papers \cite{voss,b1,b2,b3,calvo,pisarchik,pyragas,fernanda}, anticipated synchronization has been experimentally observed in electronic circuits \cite{b2,vossexp,pisarchikexp} and laser systems \cite{liu1,liu2} and it has been proposed as a mechanism to control dynamical systems \cite{xu,marzena09,qin}, competing with more traditional techniques as the Ott, Grebogi and York \cite{ott} or the Pyragas \cite{pyragas92} methods. The idea behind the control technique using anticipated synchronization, named as the {\sl predict-prevent} control method \cite{marzena09}, lies in the use of the information obtained from the slave output to prevent unwanted behaviors in the master system.

In this paper we perform an extensive numerical study of the predict-prevent control method for excitability using FitzHugh-Nagumo model systems \cite{koch,glass} as prototypic examples. We show that it is possible to control the master system by monitoring the output of the slave by using this method. We study the robustness of the control for different types of external forcing as well as with respect to the parameter mismatch between master and slave.

Although anticipated synchronization has been described before for excitable systems \cite{b1,b2,masoller,b4}, a detailed characterization of the anticipated synchronization region in parameter space and its dependence with the forcing is still lacking, specially when using different forcing terms in the master and the slave. Therefore, we devote section \ref{sec:equations} to the description of the model equations as well as the characterization of the anticipated synchronization regions. In section \ref{sec:control} we show how to control the master system by using the output of the slave system. In section \ref{sec:mismatch} we analyze the influence of parameter mismatches. Finally, in section \ref{sec:conclusions} we present the summary and conclusions.

\section{Equations and forcing schemes}
\label{sec:equations}
We consider two FitzHugh-Nagumo systems in the presence of external forcing terms and unidirectionally coupled in a master-slave configuration. The (dimensionless) equations are \cite{b1}
\begin{eqnarray}
\dot{x}_1(t)&=&-x_1(x_1-a)(x_1-1)-x_2+I_1(t),\label{eq:master1}\\
\dot{x}_2(t)&=&\epsilon (x_1-bx_2),\label{eq:master2}\\
\dot{y}_1(t)&=&-y_1(y_1-a)(y_1-1)-y_2+I_2(t)+\kappa [x_1(t)-y_1(t-\tau)],\label{eq:slave1}\\
\dot{y}_2(t)&=& \epsilon(y_1-by_2),\label{eq:slave2}
\end{eqnarray}
where $(x_1, x_2)$ are the dynamical variables of the master system and $(y_1, y_2)$ are the corresponding ones for the slave system; $a=0.139$, $b=2.54$ and $\epsilon=0.008$ are constant parameters (as $\epsilon\ll 1$ there is a separation of time scales between the fast, $x_1, y_1$, and the slow, $x_2, y_2$, variables). $I_1(t)$ and $I_2(t)$ represent external forcings. In Eq. (\ref{eq:slave1}) the term $\kappa x_1(t)$ represents the unidirectional coupling from the master to the slave, while $-\kappa y_1(t-\tau)$ is a feedback term delayed in time by an amount $\tau$ ($\kappa$ controls the strength of both the coupling and the feedback). Note that variables $x_1$ and $y_1$ are coupled in such a way that the dynamics of the master influences, but is independent of, the dynamics of the slave.

\begin{figure}[H]
\centerline{%
\includegraphics[scale=0.55]{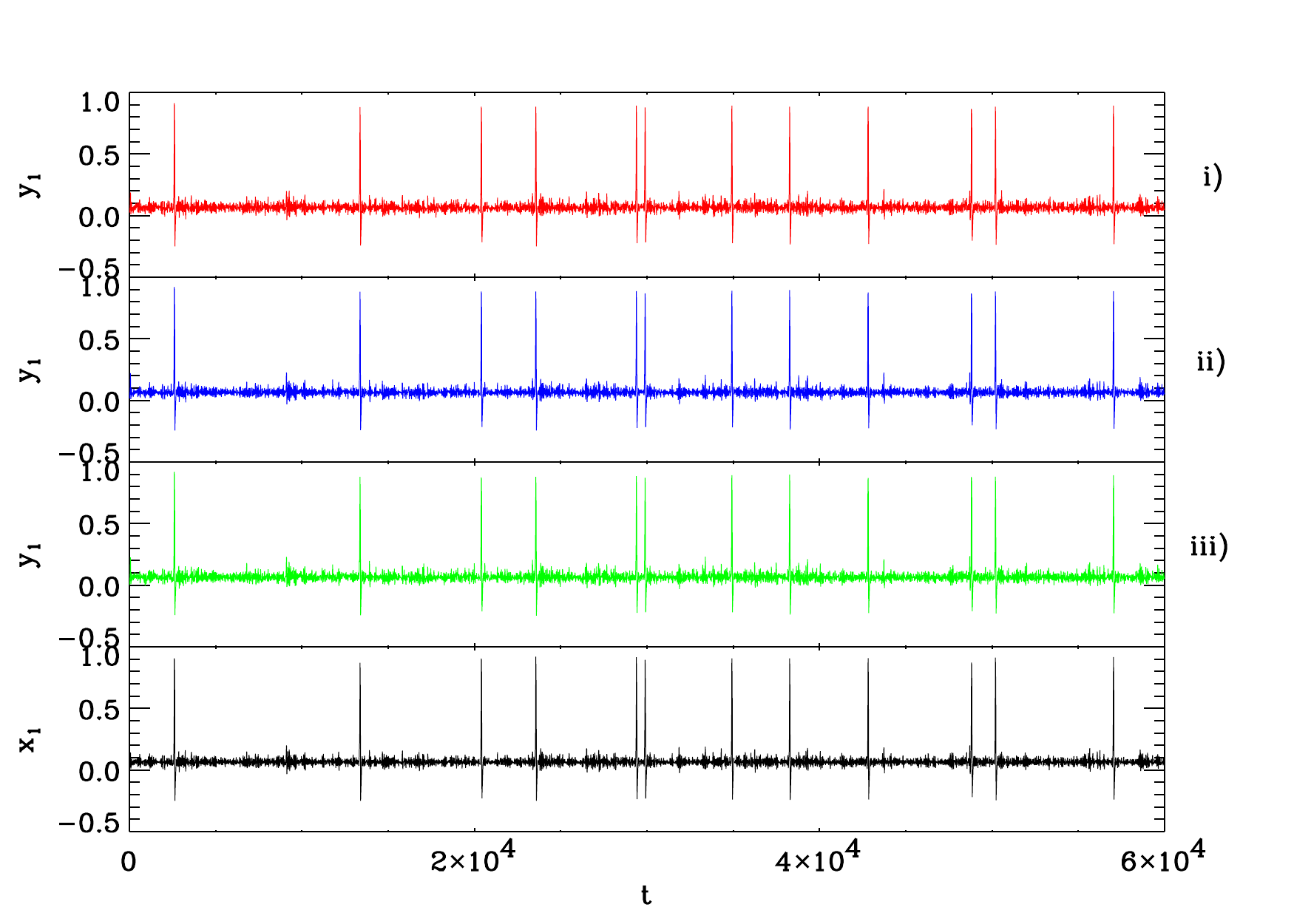}}
\caption{(Color online) Trajectories $x_1(t)$ and $y_1(t)$ coming from a numerical integration of Eqs. (\ref{eq:master1}-\ref{eq:slave2}). The forcing in the master is always $I_1(t)=I_0+\xi_i(t)$ being $\xi_1(t)$ white-noise, and the corresponding trajectory is plotted in the bottom panel (black line). We also plot the slave trajectories when different forcing schemes (i)-(ii)-(iii) (see the main text) are applied to the slave.  Parameters: $a=0.139$, $b=2.54$, $\epsilon=0.008$, $\kappa=0.4$, $\tau=2.4$, $I_0=0.03$, $D=2.45\cdot10^{-5}$.
\label{fig:timeseries}}
\end{figure}

Let us first describe briefly the main characteristics of the dynamics of the master system. Without external forcing, there exists a fixed point at the origin $x_1=x_2=0$. This fixed point is stable against small perturbations induced by the external forcing $I_1(t)$, but if the perturbations exceed a threshold value, then the system returns to the fixed point by a large excursion in phase space, a so-called {\sl spike} or {\sl pulse}. If, on the other hand, $I_1(t)=I_0$, a small constant, the stable fixed point is slightly different from zero. More specifically, for $I_0<0.035$ (for the set of parameters $a,b,\epsilon$ noted before) the fixed point is stable (and excitable behavior can occur), whereas for $I_0>0.035$ the dynamics is oscillatory. Of course, the case of a constant forcing is not very interesting and we will consider that the forcing acting on the master system can be decomposed as $I_1(t)=I_0+\xi_1(t)$, being $\xi_1(t)$ a random function of time and we take throughout the paper $I_0=0.03$, below the excitability threshold. We have considered two possibilities: (1) $\xi_1(t)$ is white noise with zero mean and correlations $\langle \xi_1(t)\xi_1(t')\rangle=D\delta(t-t')$; (2) $\xi_1(t)=D\sum_k\delta(t-t_k)$ is a sum of impulses at random times $t_k$ such that the time differences $t_{k+1}-t_k$ are distributed according to an exponential distribution of mean value $\lambda$. In both cases we call $D$ the noise intensity. As a consequence of the random forcing, the system displays pulses at random times, see Fig. \ref{fig:timeseries}.

Let us now consider the dynamics of the slave system. For constant common forcing ($I_1(t)=I_2(t)=I_0$), there is a solution of the previous equations in which $y_1(t)=x_1(t+\tau)$ and $y_2(t)=x_2(t+\tau)$. This remarkable solution, first found by Voss \cite{voss}, shows that the slave anticipates (i.e. predicts) by an amount of time $\tau$ the dynamics of the master. Our intention is to use this anticipation property of the slave to influence the dynamics of the master in order to suppress all unwanted pulses.

\begin{figure}[H]
\centerline{
\includegraphics[scale=0.55]{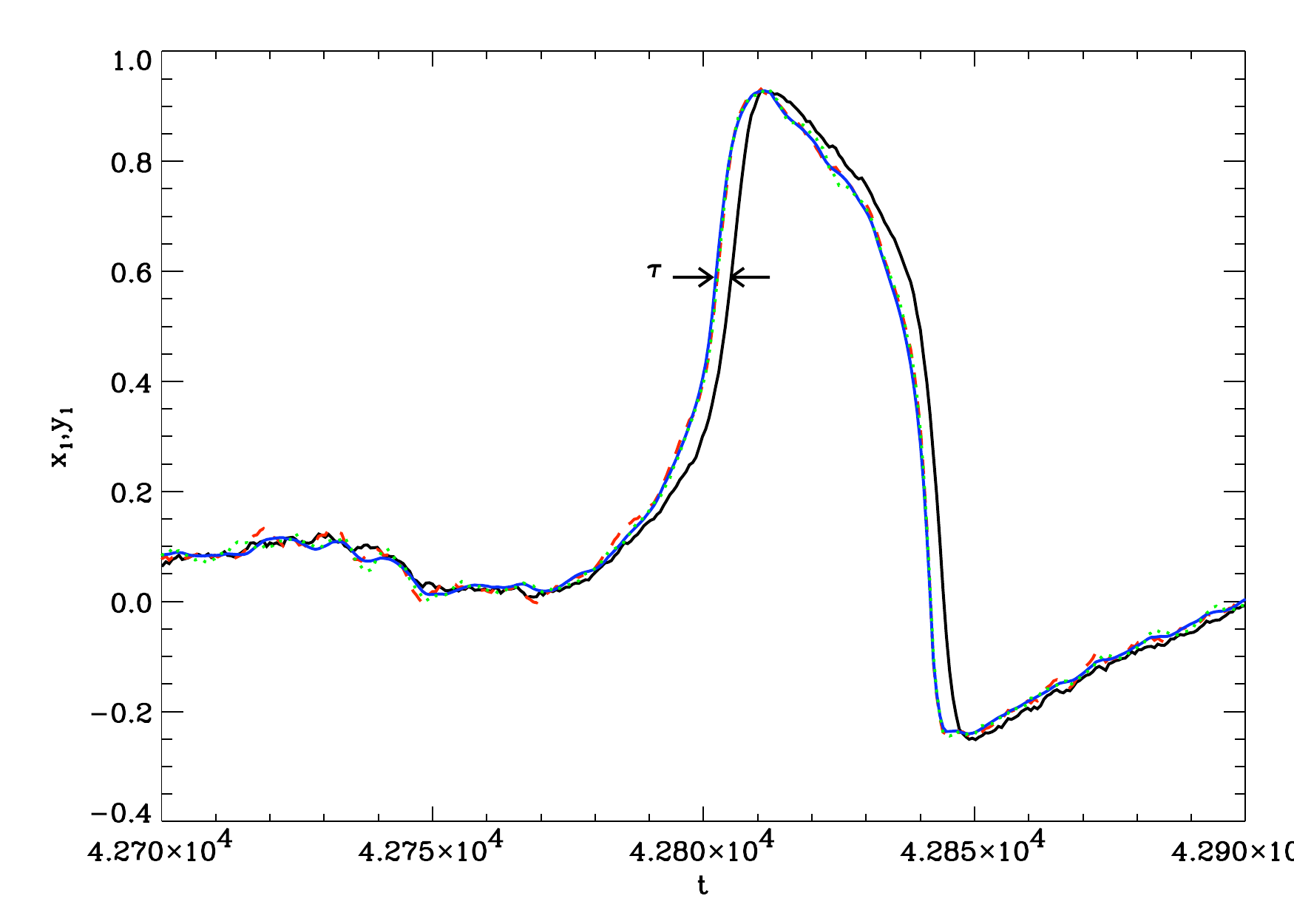}}
\caption{(Color online) Blow up of Fig. \ref{fig:timeseries} to show the details of a single pulse. Note that, in all cases (i)-(ii)-(iii) the slave anticipates the master by a time approximately equal to $\tau$.
\label{fig:pulse}}
\end{figure}

For non-constant $I_1(t)$ it is no longer true that the anticipated manifold is an exact solution even in the case of common forcing. We will consider that the forcing on the slave is either constant, $I_2(t)=I_0$, or it can be decomposed as $I_2(t)=I_0+\xi_2(t)$ with $\xi_2(t)$ a white noise of intensity $D$. In some cases we will take $\xi_2(t)=\xi_1(t)$ and in others that $\xi_2(t)$ and $\xi_1(t)$ are independent random processes. We have shown in previous work \cite{b2} that, in this more general forcing scenario, it can exist nevertheless a region in the parameter space $(\kappa,\tau)$ such that it holds that $y_1(t)\approx x_1(t+\tau)$ and, in particular, that the slave can display pulses that anticipate by a time approximately equal to $\tau$ the pulses of the master. In the following, we analyze the details of this anticipated synchronization region in four cases:\\
 {\bf (i)} identical random white noises $\xi_1(t)=\xi_2(t)$,\\  {\bf (ii)} white noise $\xi_1(t)$ in the master and constant forcing $I_0$ in the slave, \\ {\bf (iii)} independent white noises $\xi_1(t)$ and $\xi_2(t)$, and \\ {\bf (iv)} a sum of impulses for $\xi_1(t)$ in the master and a constant forcing $I_0$ in the slave. \\ In each case, we fix the values of $I_0$ and $D$ and study the region of parameter space $(\kappa,\tau)$ in which the slave anticipates correctly the pulses of the master. In Fig. \ref{fig:timeseries} we show an example of a trajectory and a detail of the anticipated synchronization in Fig. \ref{fig:pulse}.

\begin{figure}[H]
\centerline{
\includegraphics[scale=0.65]{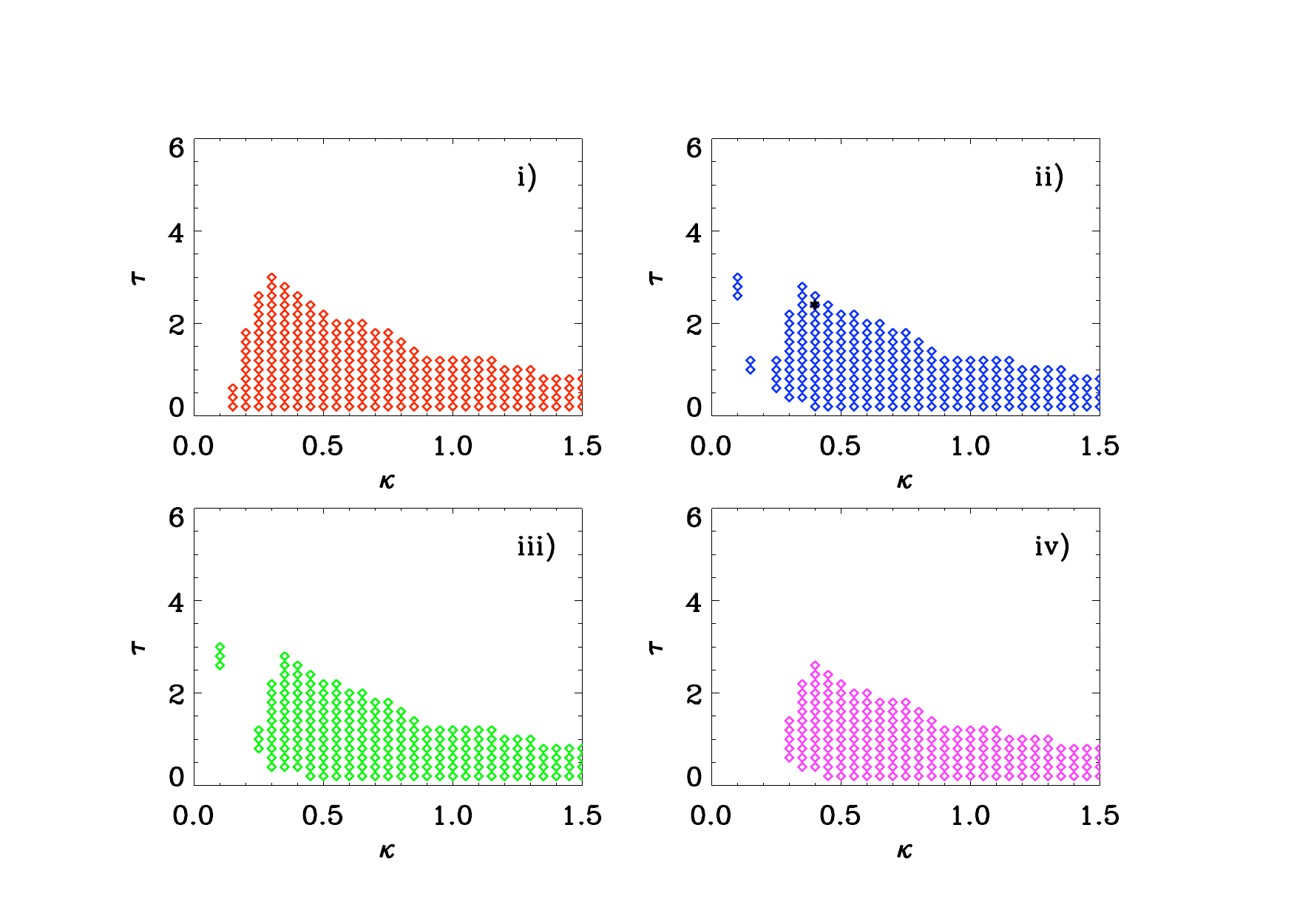}}
\caption{(Color online) The dots show the anticipated synchronization region defined as the set of values of $(\kappa,\tau)$ for which $\rho_1<0.1$, $\rho_2<0.1$ and $\rho_3<0.2$ in the four forcing cases (i)-(ii)-(iii)-(iv) explained in the main text. Other parameter values as in Fig. \ref{fig:timeseries} except in (iv): $I_0=0.032$, $D=0.05$, $\lambda=500$. 
\label{fig:region}}
\end{figure}

The region in parameter space in which anticipation of pulses is possible is shown in Fig. \ref{fig:region} for the different forcings (i)-(ii)-(iii)-(iv) described before. In each case we have set fixed values of $I_0$, $D$ and quantified the quality of the anticipated synchronization by the following measure: $\rho_1=|n(x_1) - n_c(y_1)|/n(x_1)$ being $n(x_1)$  the number of pulses in the master system observed in a given (large) time interval and $n_c(y_1)$ the number of those pulses which are correctly predicted by the slave system. We have considered that a pulse at a time $t$ in the master is correctly predicted if there is another pulse in the slave occurring at a time $t'$ ($t$ (resp. $t'$) are taken when the master (resp. slave) crosses the value $0.6$) such that $\Delta t \equiv t-t'$ satisfies $0<\Delta t<20$ (this time being about half the width of a typical pulse). As it turns out that the slave might display some extra, spurious, pulses with no correspondence in the master, it is also necessary to introduce $\rho_2=|n(y_1) - n(x_1)|/n(x_1)$, being $n(y_1)$ the number of pulses observed in the same time interval in the slave system. Furthermore, as we are looking for a real anticipation between the pulses, we need to know that the average anticipation time is close to $\tau$ and, hence, we introduce the third measure $\rho_3=|\langle\Delta t\rangle-\tau|/\tau$. Perfect anticipated synchronization implies $\rho_1=\rho_2=\rho_3=0$. The anticipated synchronization region in Fig. \ref{fig:region} signals the values of $(\kappa,\tau)$ for which $\rho_1<0.1$, $\rho_2<0.1$ and $\rho_3<0.2$. In agreement with the results of reference \cite{marzena09}, in which case (i) was considered using a different measure of synchronization, we find that for too large or too small coupling $\kappa$ anticipation is lost. Moreover, a too large delay time $\tau$ also prevents anticipated synchronization of pulses to occur.
In the next section, we will consider parameter values in which anticipation does occur and we will devise a method that will allow us to suppress in an efficient way the pulses in the master system.

\section{Control}
\label{sec:control}
As stated before, our goal is to suppress the appearance of the random pulses in the master system, consequence of an unavoidable random forcing $I_1(t)$. To this end, we consider first a simple direct control procedure that consists in reducing the amplitude of the $x_1(t)$ variable whenever it surpasses a threshold value $x_0>0$. This simple scheme is such that if $t_0$ is the time at which $x_1(t_0)=x_0$ crosses the threshold value, then we set $x_1(t=t_0^+)=x_1(t_0)-p$, being $p$ the amplitude of the control `kick'. This is equivalent to adding a term of the form $-p\delta(t-t_0)$ to Eq. (\ref{eq:master1}). In this direct control method, we have to be careful about the precise value of $x_0$. If $x_0\approx 0$ then the condition $x(t)=x_0$ might not lead necessarily to a pulse and we would be applying a control condition when it is not needed. If, on the other hand, $x_0$ is very large, the pulse will be already too developed and it will be necessary to apply a strong control amplitude $p$ in order to suppress it.

In the following, we consider a modification of this simple direct control method. Following the ideas developed in reference \cite{marzena09}, we show that the use of a predict-prevent control scheme, based upon the slave variable crossing the threshold value, $y_1(t_0')=x_0$, is more efficient in the sense that it allows to suppress more pulses with a smaller control amplitude $p$. Likewise, the use of a value of $p$ as small as possible is important in order not to introduce an uncontrolled perturbation to the dynamics of the master, as large values of $p$ might induce the appearance of additional pulses after the suppressed ones.

\begin{figure}[h]
\centering
\includegraphics[scale=0.55,angle=90]{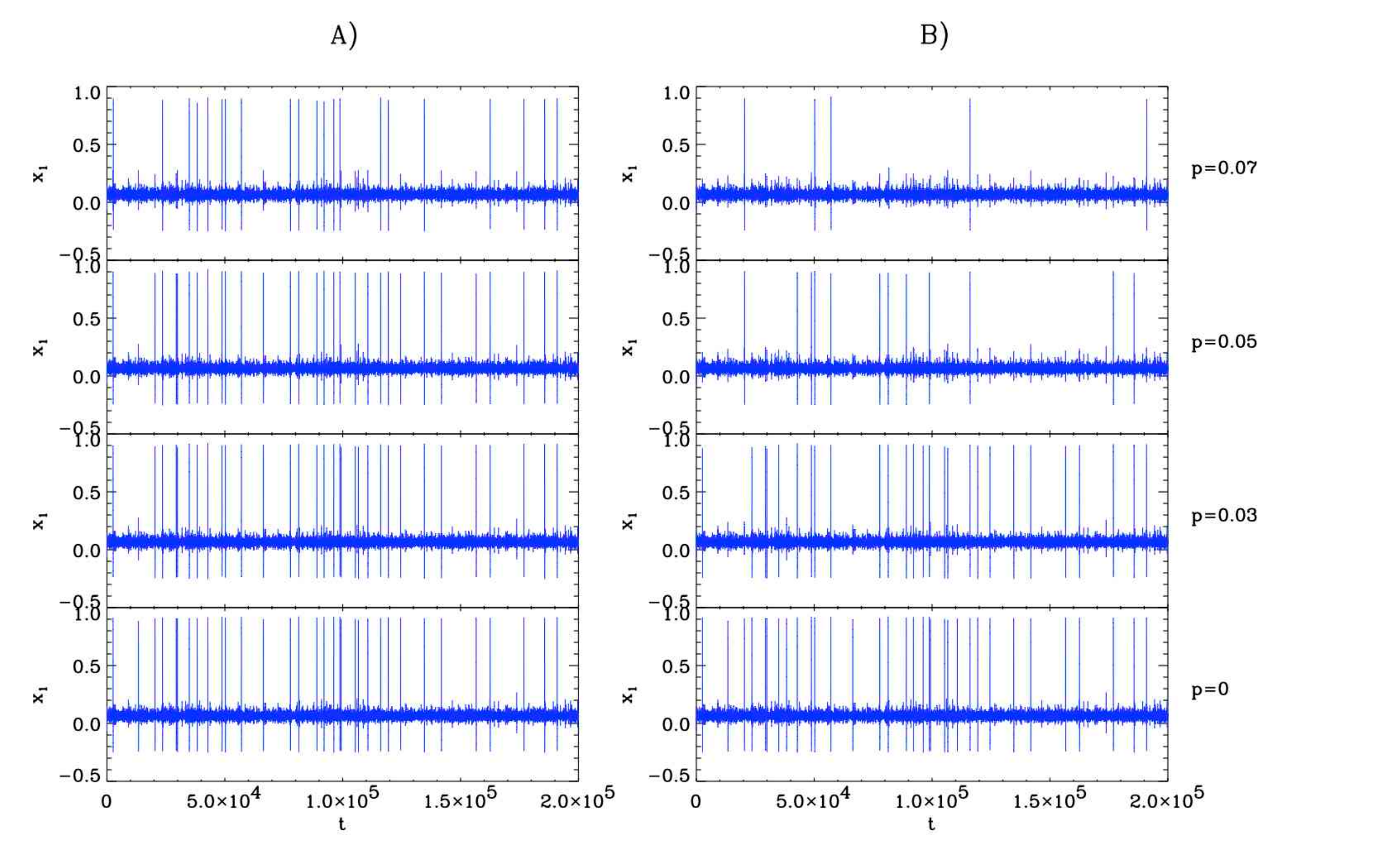}
\caption{(Color online) Time traces of the master variable $x_1(t)$. In panels A, left, we show the pulse suppression after applying a direct control (`kick') of magnitude $p$ whenever the master variable $x_1(t)$ crosses the threshold value $x_0=0.3$. As $p$ increases the number of pulses successfully suppressed increases with respect to the no-control case, $p=0$. In the right panels, B, we show that the reduction is more effective if we use a control scheme based upon the slave variable $y_1(t)$ crossing the same threshold value. The results come from a numerical integration of Eqs. (\ref{eq:master1}-\ref{eq:slave2}) in the forcing scheme (ii): a random forcing $I_1(t)=I_0+\xi_1(t)$, being $\xi_1(t)$ Gaussian noise of intensity $D$, in the master and a constant forcing $I_2(t)=I_0$ in the slave. Parameters are: $a=0.139$, $b=2.54$, $\epsilon=0.008$, $I_0=0.03$, $D=2.45\cdot10^{-5}$, $\kappa=0.4$, $\tau=2.4$ ($\kappa$ and $\tau$ correspond to the black dot in Fig. \ref{fig:region}). The recovery time after which no other correction can be applied is $t_{rec}=100$.
\label{fig:control}}
\end{figure}

In Fig. \ref{fig:control} we plot the resulting time series for the variable $x_1(t)$ after applying the control procedure just described in case (ii): white noise applied to the master and a constant forcing to the slave. It can be clearly seen that, for the same threshold value $x_0$ and control amplitude $p$, a more efficient control procedure (i.e. a larger fraction of suppressed pulses) occurs when using the predict-prevent control scheme based upon the slave variable than the one based on the direct control method of the master. The reason is obvious, as the pulses of $y_1(t)$ precede by a time approximately equal to $\tau$ the corresponding pulses of $x_1(t)$, the control is applied earlier in time, when the pulse in the $x_1(t)$ is not so well developed yet and it is easier to suppress. In all cases, and in order to avoid spurious and repeated control kicks, we have set a recovery time $t_{rec}=100$ (this is a value larger than the average time-width of a pulse), such that two consecutive correcting kicks can not be applied in a time shorter that $t_{rec}$.

\begin{figure}[H]
\centerline{
\includegraphics[scale=0.55]{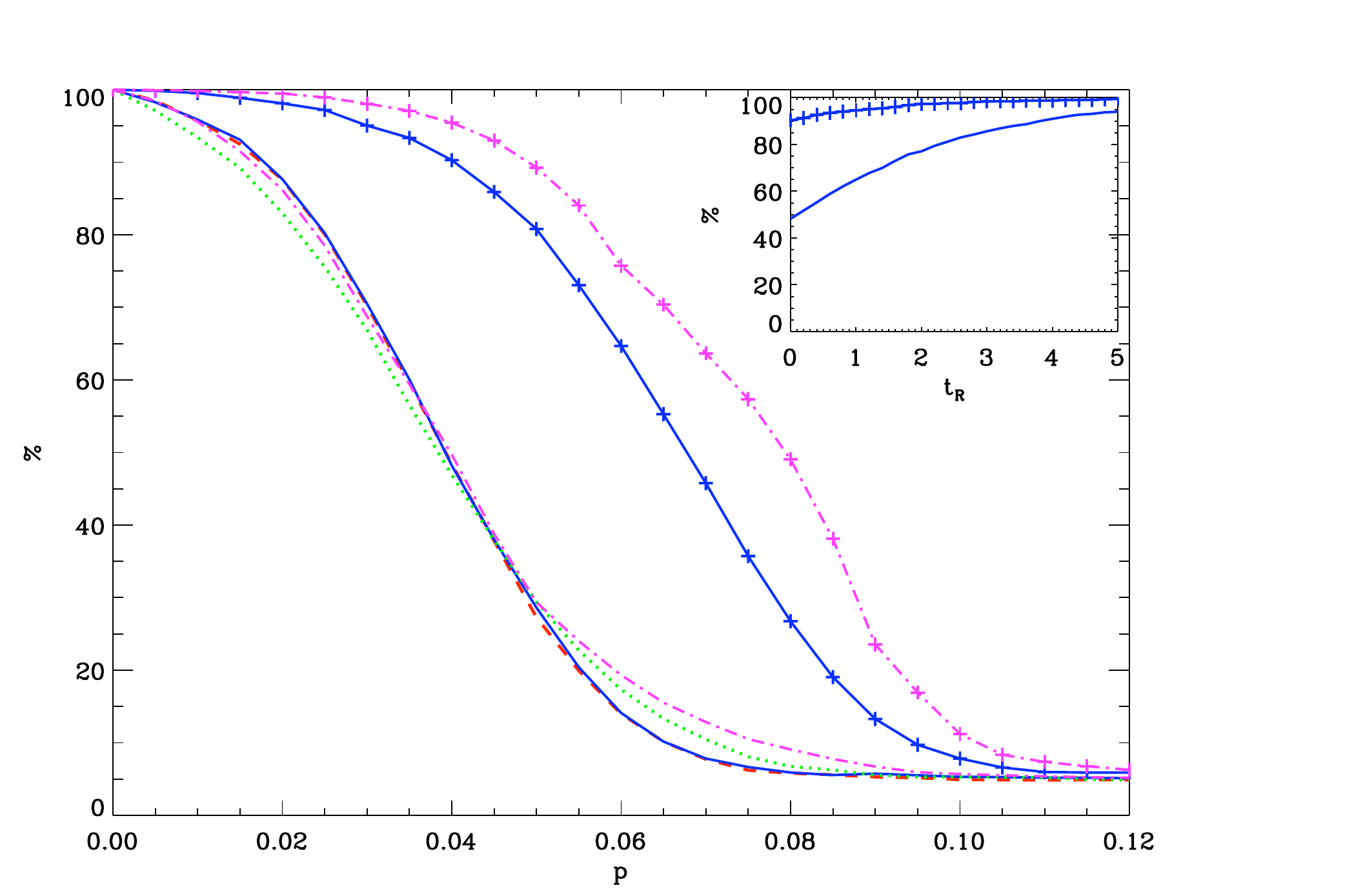}}
\caption{(Color online) Fraction of remaining pulses of $x_1$ as a function of the control parameter $p$. The lines with + symbols correspond to the direct control scheme based only upon the master system for forcing schemes (i)-(ii)-(iii) (solid) and (iv) (dot-dashed). Same parameter values than in Figs. \ref{fig:region} and \ref{fig:control}. The different lines at the left of the main graph correspond to the predict-prevent control using the slave in the different forcing schemes: (i), red dashed; (ii), blue solid; (iii) green dotted; and (iv),  purple dot-dashed. In the inset, we show the effect of a reaction time $t_R$ that elapses between the control variable crosses the threshold value $x_0$ and the corrective impulse is applied for case (ii) and $p=0.04$ using direct control on the master (solid line with $+$) or predict-prevent control on the slave (solid line).
\label{fig:error}}
\end{figure}

In Fig. \ref{fig:error} we quantify the error of the control scheme by plotting the fraction of pulses that were not successfully suppressed as a function of the control amplitude $p$. As mentioned above, for large $p$ the control procedure might lead to the appearance of new pulses, but, in any event, we see that the number of remaining pulses is a decreasing function of $p$. This figure shows that a substantial improvement in the suppression of pulses is obtained when using the predict-prevent control method based on the slave system. This is the main result of this paper.

To show the robustness of the control scheme based on the slave system, and in order to cover a wider range of possible experimental situations, we show in the same figure the fraction of not suppressed pulses for the whole set of forcing schemes (i)-(ii)-(iii)-(iv) defined before. In all cases, a much better reduction with respect to the direct control using only the master is achieved.

We have also studied the effect of a time lag $t_R$ between the crossing of the signal control with $x_0$ and the application of the control at $x_1$. As it can be seen in the inset of Fig. \ref{fig:error} for forcing case (ii), as $t_R$ increases, less pulses are suppressed, both in the direct and the predict-prevent control procedures. These results imply that for a successful pulse suppression the time between the crossing of the signal and the application of the control must be as short as possible. We have observed similar results for the other forcing cases.

\section{Parameter mismatch}
\label{sec:mismatch}

As it is very unlikely that one can produce a perfect copy of the master system, an important issue concerning the control based on a master-slave configuration is how robust it is upon differences in parameter mismatch between the two systems. In fact, an experiment using an electronic implementation of the FitzHugh-Nagumo neurons has been carried out in reference \cite{marzena09} with the result that the control procedure can be carried out safely with real, non-identical, systems.

We have studied the effect of small variations of the parameters $a$, $b$ and $\epsilon$ in the slave system, Eqs. (\ref{eq:slave1}-{\ref{eq:slave2}}), in order to analyze how the control of the master dynamics is affected. We have first studied the effect of each one of these parameters separately by changing about $5\%$ its value. As a consequence the fraction of remaining pulses changes by: $20\%$ in the case of varying $a$, $10\%$ in the case of varying $b$, and $8\%$ in the case of varying $\epsilon$. For values of $a$ in the slave larger than  in the master less pulses are controlled. A similar result applies to variations of $\epsilon$. However, the effect of $b$ is different: for values of $b$ larger in the slave than in the master, the fraction of remaining pulses decreases when using the predict-prevent control method, i.e, more pulses are controlled.
The effect of $a$ and $b$ can be understood in terms of the stable fixed point: For different parameters in the slave and the master, the stable fixed points will be different. For larger values of $a$ in the slave system, the fixed point moves towards the origin, so decreasing the excitability threshold. In contrast, for $b$ larger in the slave the fixed point moves away from the origin, so increasing the excitability threshold and worsening the pulse control.

\begin{figure}[H]
\centerline{
\includegraphics[scale=0.55]{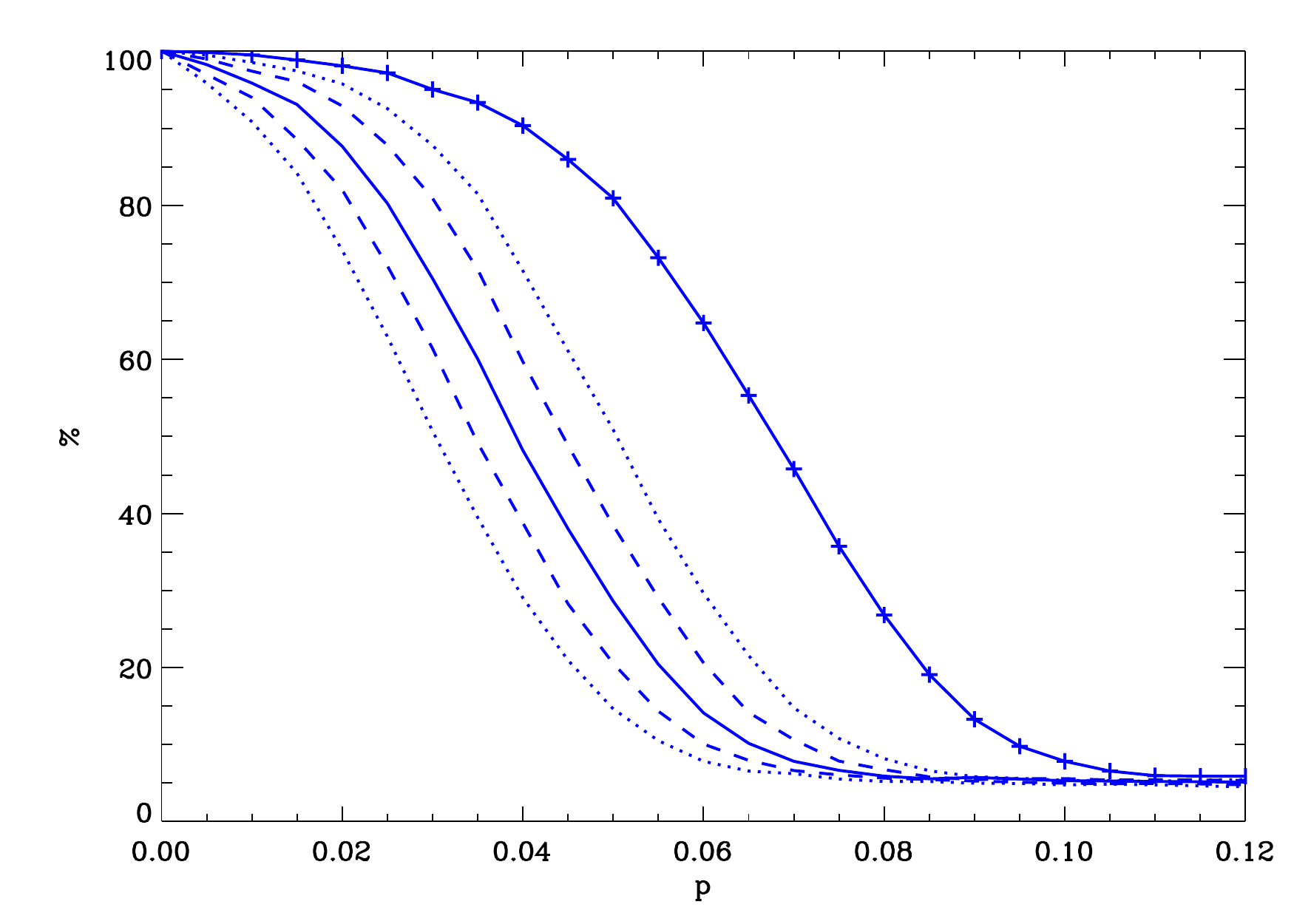}}
\caption{(Color online)  In order to show the effect of parameter mismatch, we plot the fraction of remaining pulses of $x_1$ as a function of $p$ in forcing case (ii) using the predict-prevent control method. The solid line corresponds to identical parameters in master and slave: $(a,b,\epsilon)=(0.139,2.54,0.008)$. The values between the dotted (resp. dashed) lines correspond to the parameters in the slave varying up to 10\% (resp. 5\%) with respect to those of the master. The solid line with + symbols is the same than in Fig. \ref{fig:error} and it corresponds to the direct control method and is plotted here for comparison. Other parameters as in Figs. \ref{fig:region} and \ref{fig:control}.
\label{fig:mismatch}}
\end{figure}

In Fig. $\ref{fig:mismatch}$, the solid lines are the same than the ones shown in Fig. \ref{fig:error}, where $a$, $b$ and $\epsilon$ in the slave equations are equal to those of the master equation.
We have studied how the control dynamics is affected when changing at the same time the three parameters ($a$, $b$, $\epsilon$) in the slave equations. For up to a $5\%$ (resp. $10\%$) of difference between the parameters of the master and the slave, the fraction of pulses that were not successfully controlled for $x_1$ are located between the two dashed (resp. dotted) lines. An specific example is to take the following parameters for the slave system: $(a,b,\epsilon)=(0.1251,2.794,0.0072)$, i.e.  $a$ smaller, $b$ larger and $\epsilon$ smaller in the slave than in the master. In this case, less pulses remain, so making the control procedure more effective (dotted line at the left of Fig. \ref{fig:mismatch}). This shows that parameter mismatch does not necessarily worsen the results, as in some cases even a larger fraction of pulses can be successfully controlled.
It is worth noting that we have observed similar results for the different forcing types that we have considered.

\section{Conclusions}
\label{sec:conclusions}

In this paper we have first characterized the anticipated synchronization region of two unidirectionally coupled Fitzhugh-Nagumo systems in a master-slave configuration for different types of forcing terms, obtaining qualitatively similar regions for the different cases (i)-(ii)-(iii)-(iv) considered. This implies that anticipated synchronization is indeed a robust phenomenon even under different types of uncommon forcing for the master and the slave equations.

Then we have performed a numerical study of the predict-prevent control method. For parameter values inside the anticipated synchronization region, we have applied two control schemes in order to suppress the pulses of the master: the direct control using the master output and the predict-prevent control using the slave output. We have obtained that the predict-prevent control is more efficient than the direct control using only the master. This statement is true for a variety of random forcing terms, both common and uncommon to the master and the slave.

Finally, we have obtained that a perfect matching between the parameters of the master and the slave is not necessary for the control to be efficient and, in fact, a slight  parameter mismatch can, in some cases, lead to a better control.

The results obtained in this work are a clear indication of the robustness of the proposed predict-prevent control method and opens the door to more general experimental implementations, in other physical and biological systems, than the ones carried out previously\cite{marzena09}.

\section*{Acknowledgments}
We acknowledge the financial support of project FIS2007-60327 from MICINN (Spain) and FEDER (EU).

\newpage

\newpage

\end{document}